\documentclass{article}
\usepackage{spconf}
\usepackage{amsmath}
\usepackage{dsfont}
\usepackage{graphicx, graphics, theorem, times, amsfonts, amsmath, amssymb, cite}
\usepackage{tikz}
\usetikzlibrary{shapes,arrows}
\usepackage{pgfplots}
\usepackage{color}
\usepackage{hyperref}
\usepackage{multirow}
\usepackage{rotating}
\usepackage{cleveref, subcaption}
\usepackage{bm}
\input{mysymbol.sty}
\usepackage[utf8]{inputenc}
\usepackage{enumitem}
\usepackage[linesnumbered]{algorithm2e}
\usepackage{array}
\newcolumntype{P}[1]{>{\centering\arraybackslash}p{#1}}

\newcommand{\QED}{\hfill\ensuremath{\blacksquare}}

\newtheorem{remark}{\hspace{-1pt}\bf Remark}

\ninept


     \makeatother

     \makeatletter
     \renewcommand\small{%
     \@setfontsize\small\@ixpt{11}%
     \abovedisplayskip 3.5\p@ \@plus3\p@ \@minus2\p@
     \abovedisplayshortskip \z@ \@plus2\p@
     \belowdisplayshortskip 3\p@ \@plus2\p@ \@minus2\p@
     \def\@listi{\leftmargin\leftmargini
     \topsep 4\p@ \@plus2\p@ \@minus2\p@
     \parsep 2\p@ \@plus\p@
     \@minus\p@
     \itemsep
     \parsep}%
     \belowdisplayskip
     \abovedisplayskip
     }
     \makeatother

\title{Learning Graphs and Simplicial Complexes from Data}
\name{Andrei Buciulea$^{\dagger}$ \qquad  Elvin Isufi$^\ddagger$ \qquad Geert Leus$^\ddagger$ \qquad Antonio G. Marques$^\dagger$
\thanks{Work supported by the EU H2020 Grant Tailor (No 952215); the Spanish NSF Grants (MCIN/AEI/10.13039/501100011033) Grants PID2019-105032GB-I00 and PID2022-136887NB-I00; the Community of Madrid (CAM) and Rey Juan Carlos University (URJC) via the Young Researchers R\&D Project ref. F861 (AUTO-BA-GRAPH), and the fellowship PREDOC20-003. Email contact author: antonio.garcia.marques@urjc.es.}}
\address{$^\dagger$Dept. of Signal Theory and Communications, King Juan Carlos University, Madrid, Spain \\ $^\ddagger$Delft University of Technology, Delft, The Netherlands}

\begin{document}
\ninept
\maketitle
\begin{abstract}
\vspace{0.4cm}
Graphs are widely used to represent complex information and signal domains with irregular support. Typically, the underlying graph topology is unknown and must be estimated from the available data. Common approaches assume pairwise node interactions and infer the graph topology based on this premise. In contrast, our novel method not only unveils the graph topology but also identifies three-node interactions, referred to in the literature as second-order simplicial complexes (SCs). We model signals using a graph autoregressive Volterra framework, enhancing it with structured graph Volterra kernels to learn SCs. We propose a mathematical formulation for graph and SC inference, solving it through convex optimization involving group norms and mask matrices. Experimental results on synthetic and real-world data showcase a superior performance for our approach compared to existing methods.

\end{abstract}
\section{Introduction}\label{sec:introduction}



Estimating the topology of complex data is a crucial step in the downstream signal processing and machine learning tasks ~\cite{mateos2019connecting,dong_2019_learning}. To estimate this structure, it is essential to model the coupling between the topology and the data and how they influence each other. For example, graph topology inference methods assume that pair-wise node-to-node interactions could explain the data behavior or their dynamics. These approaches rely on algebraic and statistical methods to infer the graph topology from the observed data. Classic examples include correlation-based methods~\cite[Ch. 7.3.1]{kolaczyk2009book}, graphical lasso (GL)~\cite{GLasso2008}, and GSP based models, which exploit signal properties such as smoothness or graph stationarity ~\cite{Kalofolias2016inference_smoothAISTATS16,egilmez2017graph,segarra2017network,buciulea2022topoid}.


Although pairwise interactions reveal some of the intricate dependencies and dynamics inherent in networked data, many interactions within groups comprise more than two nodes \cite{net_motifs,bick2023higher}. For example, research collaborations often involve teams of authors and molecules tend to interact in small groups rather than pairs. To model such group interactions, common approaches resort to hypergraphs \cite{hypergraphs,young2021hypergraph} or simplicial complexes\cite{hodge_lap, sc_barbarosa, yang2022simplicial,SCHAUB2021edge}. The latter are typically either considered as given or estimated via simple domain-specific heuristics. The work in \cite{dualsmooth} estimates hypergraphs from data by assuming a smoothness behavior on node and edge signals. It first infers a graph topology and then constructs on it a line graph to retrieve the higher-order interactions (hyperedges) but does not directly reveal the latter. This underscores the need for a model that is capable of learning jointly the graph structure and capturing higher-order relationships between nodes.
To account for the latter, we resort to Volterra models on networks, which model node dynamics in a nonlinear manner involving both pair-wise and higher-order interactions \cite{yang2023autoregressive, leus2021topological,H-Xli2009volterra}.

Specifically, we consider a networked autoregressive Volterra model and pose an inverse problem to \emph{jointly} estimate the graph and the higher-order connectivity only from node signal realizations. Higher-order connectivities are embedded in the Volterra kernels of such a model. To limit the degrees of freedom (DoFs), we impose an SC structure between node-to-tuple interactions, ultimately, establishing a relationship between SCs and Volterra kernels. For the particular case of an SC of order two (representing connectivities up to triplets), this relationship simplifies to relating the Volterra kernel of order one to graph edges (simplex of order one) and the Volterra kernel or order two to filled triangles (simplex of order two). Subsequently, we develop a convex formulation that incorporates group sparsity to solve the proposed problem. The group sparsity allows us to group edges and node-to-tuple interactions related to each triplet of nodes and to control the number of triplet interactions. The proposed approach is corroborated via numerical experiments on synthetic and real data showing a competitive performance compared with alternatives that estimate either the graph topology or rely on it to infer higher-order interactions.



\section{Problem Formulation}\label{S:netw_reconstruction}

Consider $R$ observations of a vector $\bbx \in \reals^N$ grouped in matrix $\bbX := [\bbx_1, \ldots, \bbx_R] \in \reals^{N \times R}$. The data entries in $\bbx$ have a hidden underlying structure, which is typically represented through a graph $\ccalG = \{\ccalV, \ccalE\}$ comprising a set of nodes $\ccalV = \{1, \ldots, N\}$ and a set of edges $\ccalE = \{(i,j) | i,j \in \ccalV\}$. Here, the $i$th entry $x_i$ is seen as a datum associated with node $i$, hence, set $\ccalE$ represents pairwise dependencies between the data of nodes $i$ and $j$. In this context, we also refer to vector $\bbx$ as the graph signal. Estimating the graph topology from the data $\bbX$ boils down to leveraging a model that expresses this data coupling and the role of the topology in it; i.e., solving an inverse problem of the form $\ccalG = f^{-1}(\bbX)$ where $f(\cdot)$ is a function acting upon the graph $\ccalG$ and modeling how it is coupled with the signal realizations in $\bbX$. Typically, we will estimate an algebraic representation of graph $\ccalG$ that is represented by its adjacency matrix $\bbA\in\reals^{N \times N}$, graph Laplacian $\bbL:= \diag(\bbA \textbf{1}) - \bbA$, or more generally a graph shift operator (GSO) matrix  $\bbS \in \mathbb{R}^{N \times N}$, where $S_{ij} \neq 0$ if and only if $i=j$ or $(j,i) \in \ccalE$ \cite{ortega_2018_graph}. The existing literature provides different approaches to estimating the graph from the data, with different assumptions on the function $f(\cdot)$ and topological assumptions such as directed, weighted or undirected edges on the sought graph. Moreover, since their goal is to recover a graph they focus on pairwise interactions.

We consider that the graph signal is influenced by the signal values in the other nodes both via pairwise and higher-order interactions. Specifically, we model the dependencies via an autoregressive graph \emph{Volterra} model of second order of the form
\begin{subequations}\label{E:x1model2_commonlabel_Volterra}
\begin{alignat}{2}\label{E:x1model2}
    \bbX &= \bbH_1\bbX + \bbH_2\bbY + \bbV + \bbE,\\
    &\text{with~}\bbY=\bbX\odot \bbX.\label{E:x1model2b}
\end{alignat}
\end{subequations}

Here, $\bbH_1 \in \reals^{N\times N}$ represents the pairwise interactions and the term $\bbH_1\bbX$ captures the part of the graph signal that can be represented as a linear combination of the signals in the other nodes. Instead, matrix $\bbH_2 \in \reals^{N \times N^2}$ is a node-to-tuple interaction matrix representing higher-order interactions between a node $k$ and a tuple $(i,j)$ in its entry $\bbH_2[k,(i,j)]$ \footnote{With a slight abuse of notation, we will alternate between ${\bf H}_2[k,(i,j)]$ and ${\bf H}_2[k,(i-1)N+j]$ to denote the value of the $k$-th row and $((i-1)N+j)$-th column of an $N\times N^2$ matrix.}.
Matrix $\bbY = \bbX \odot \bbX \in \reals^{N^2 \times R}$ is obtained by performing the Khatri-Rao product (column-wise Kronecker product) on the graph signals. The $r$-th column of $\bbY$ collects all the monomials of degree two involving variables $\{x_r^i\}_{i=1}^N$. These product signals can be seen now as values associated with tuples of nodes. Hence, the $r$-th column of $\bbH_2\bbY$ captures the part of a graph signal ($\bbx_r$) that can be represented by $\bbx_r \odot \bbx_r$ via node-to-pair interactions ($\bbH_2$). Finally, $\bbV\in\reals^{N\times R}$ is an exogenous variable and $\bbE\in\reals^{N\times R}$ is white zero-mean noise. For didactical purposes, we focus on a single node $k$ and model \eqref{E:x1model2_commonlabel_Volterra} allows writing its signal as\footnote{To ease exposition, we assume $R = 1$ and drop the realization index.}
\vspace{-0.1cm}
\begin{align*}\label{eq.nodekupdate}
\begin{split}
        x_k = \sum_{j = 1, j \neq k}^N \!\!\!\!\!\bbH_1[{k,j}]x_j &+ \!\!\!\sum_{i = 1, i\neq k}^N\sum_{j = 1, j \neq i,k}^{N} \!\!\! \bbH_2[{k,(i-1)N+j}]x_ix_j \\ 
        &+ v_k + e_k,
\end{split}
\end{align*}
where the first and second summations are reminiscent of $\bbH_1\bbX$ and $\bbH_2\bbY$, respectively. The latter are linearly combined by the node-to-tuple weights $\bbH_2[k,(i,j)]$ here expressed as matrix entries $\bbH_2[{k,(i-1)N+j}]$. Finally, $v_k$ and $e_k$ are the exogenous variable and noise at node $k$, respectively. The following example makes this discussion more tangible.

\begin{figure}
	\includegraphics[width=0.28\textwidth]{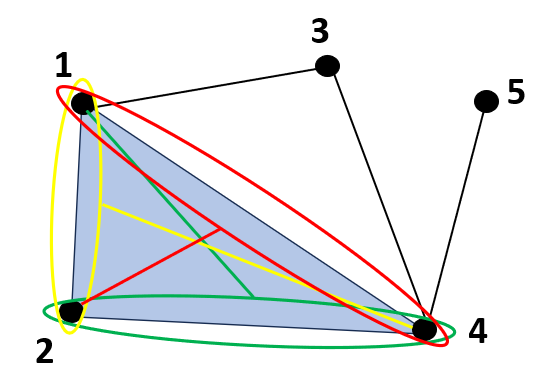}
	\centering
	\caption{The visual representation of a graph with 5 nodes and 6 edges. The node-to-pair interactions between nodes [$1$,$2$,$4$] and edges [($2$,$4$),($1$,$4$),($1$,$2$)] are represented by green, red, and yellow lines, respectively. A filled triangle between nodes $1$, $2$ and $4$ is represented in blue.}
	\label{F:high_inter}
 \vspace{-0.2cm}
\end{figure}

\smallskip
\noindent \textbf{Example.} Fig.~\ref{F:high_inter} illustrates pairwise and higher-order interactions among five nodes ($1, \ldots, 5$). The pairwise interactions (e.g., (1,2), (2,4), etc.) are shown by black solid lines. Matrix $\bbH_1$ collects them. The node signals $x_1, \ldots, x_5$ could be seen as values over the respective nodes. There are three tuples in this figure highlighted by ellipsoids; i.e., $(1,2)$ in yellow, $(1,4)$ in red, and $(2,4)$ in green. The node-to-pair interactions are shown by solid lines connecting the node to the respective tuple with the same color; i.e., [1, (2,4)], [2,(1,4)] and [4,(1,2)]. Matrix $\bbH_2$ collects all these interactions. The product node signals $x_1 x_2$, $x_1 x_4$, and $x_2 x_4$ could be seen as values associated with tuples captured by matrix $\bbY$ in \eqref{E:x1model2_commonlabel_Volterra}. Following the model \eqref{E:x1model2_commonlabel_Volterra}, the signal at node $k=2$, can be written as
\begin{align}
\begin{split}
        x_2 = \bbH_1[2,1] x_1 &+ \bbH_1[2,4] x_4 + \bbH_2[2,(1,4)] x_1 x_4 \\
        &+ \bbH_2[2,(4,1)] x_1 x_4 + v_2 + e_2.
\end{split}
\end{align}
%



Model \eqref{E:x1model2_commonlabel_Volterra} has two particular aspects worth discussing. First, it relates the graph signals to both the pairwise and higher-order connectivities in a nonlinear manner in $\bbX$ but still linear in the topological variables $\bbH_1$ and $\bbH_2$. This is reminiscent of how classical time-series Volterra models \cite{volt_series} have been extended to graph signals and proven relevant in applications such as power distribution grids, social networks, and recommender systems \cite{yang2023autoregressive, leus2021topological}, to name a few. The Volterra models have been found to provide both expressibility for higher-order interactions, as well as interpretability for further understanding of the underlying network dynamics. Second, \eqref{E:x1model2_commonlabel_Volterra} is a rather flexible backbone model that can be further enriched via more expressive kernels. For example, we could consider a higher-order Volterra model (as opposed to a second-order here) to capture node-to-triple interactions. Also, we considered monomials of order two as signals over tuples in $\bby$ but \textit{non-multiplicative} higher-order interactions between node signals might also be of interest. From this perspective, we could define $\bbY=[\bby_1,...\bby_R]$ as $\bby_r=g({\bf x}_r, {\bf x}_r)$, where $g(\cdot)$ is a general function for which the following equality holds $g(\bba,\bbb) = g(\bbb,\bba)$. 


Another advantage of the model in \eqref{E:x1model2_commonlabel_Volterra} is that it allows imposing straightforwardly a desired structure on both $\bbH_1$ and $\bbH_2$. This is particularly important if we want to estimate these matrices by a limited number of signal realizations. Particular structures include:
\vspace{-0.1cm}
\begin{enumerate}[label=\alph*)]
\item \emph{Positive weights:} This could be achieved by imposing the element-wise constraints $\bbH_1 \geq \bb0$ and $\bbH_2 \geq \bb0$.
\vspace{-0.1cm}
\item \emph{No self-loops:} This is particularly important to avoid trivial solutions, such as $\bbH_1\!\!=\!\!\bbI$ \eqref{eq.nodekupdate}. We could define two binary masking matrices $\bbB_1 \!\!\in \!\! \{0,1\}^{N\times N}$ and $\bbB_2\in\{0,1\}^{N\times N^2}$ and impose the elementwise constraints $\bbB_1\circ \bbH_1 = \bb0$ and $\bbB_2 \circ \bbH_2 = \bb0$, where $\circ$ denotes the entry-wise (Hadamard) multiplication. For example, $\bbB_1\!\!=\!\!\bbI$ implies that $\diag(\bbH_1) = \bb0$ and, as a result, removes pairwise self-loops.
\vspace{-0.1cm}
\item \emph{Symmetry:} This could be guaranteed by forcing $\bbH_1=\bbH_1^{\top}$. 
\end{enumerate}

However, this structure is mild and can still have prohibitive DoFs or lead to learned structures $\bbH_1$ and $\bbH_2$ that are too disconnected from each other. Therefore, \emph{our goal is to leverage model \eqref{E:x1model2_commonlabel_Volterra} and use nodal realizations $\bbX$ to jointly estimate the pairwise graph structure $\bbH_1$ and the higher-order connectivities $\bbH_2$ by imposing an SC structure in $\bbH_2$ that reduces the DoFs and couples the learned $\bbH_1$ with $\bbH_2$.}
\vspace{-0.3cm}
\section{Graph and Simplicial Complex Learning}\label{S:proposed_form}
\vspace{-0.2cm}
Since our goal is to learn an SC of order 2 (which can be understood as learning a graph and filled triangles), we provide a brief overview of the SC concept directly from a geometric perspective. A graph $\ccalG=\{\ccalV,\ccalE\}$, with $\ccalE\subseteq \ccalV\times\ccalV$, is an SC of order 1. An SC of order 2 can be represented by $\{\ccalV,\ccalE,\ccalT\}$, where $\ccalT\subseteq \ccalV\times\ccalV\times\ccalV$ and a triplet $(i,j,k)$ can belong to $\ccalT$ only if all pairs $(i,j)$, $(j,k)$ and $(i,k)$ belong to $\ccalE$. This readily implies that an SC of order 2 can be represented by graph $\ccalG = \{\ccalV,\ccalE\}$ together with a list of filled triangles. Similarly, for an $n$-th order SC ($n$-simplex) to exist, the presence of all $(n-1)$-th order SCs (also known as $(n-1)$-simplices) is required. This implies that higher-order interactions rely directly on the connectivity between the nodes in the graph.

To identify 2-simplices using Volterra kernels, it is essential to represent interactions among three interconnected nodes. To achieve this representation, we use the matrix $\bbH_2$, which captures the influence of the product of the signals at two nodes on a third node. 
It can be seen that, indeed, the entry $[k,(i,j)]$ of $\bbH_2$ involves three nodes. 
Conversely, the relation between nodes $k$, $i$, and $j$ appears in six elements of $\bbH_2$. 
In the context of SCs, our modeling assumption is that the six entries of $\bbH_2$ associated with the triplet $(k,i,j)$ can be different from zero only if the triangle $(k,i,j)$ is filled. 

\noindent \textbf{Example. (cont.)} Fig. \ref{F:high_inter} illustrates the correspondence between a filled triangle and node-to-pair interactions among three nodes. Nodes $(3,4,5)$ do not form a triangle (not all edges are present), and hence, the filled-triangle relation cannot exist. Differently, for the triplets $(1,3,4)$ and $(1,2,4)$ the triangle exists. Our assumption is that the triangle is filled if the node-to-edge interactions involving the three nodes (denoted as an ellipse and a straight line) exist. This is the case for nodes $(1,2,4)$ and therefore the filled triangle exists. Conversely, this condition does not hold for the triangle $(1,3,4)$, so it remains unfilled. 


Based on the previous discussion, with $\bbX[i,r]$ denoting the $(i,r)$-th entry of matrix $\bbX$, and recalling that $\bbY=\bbX\odot\bbX$ [cf. \eqref{E:x1model2b}], our approach to identifying an SC of order 2 by using an autoregressive Volterra model can be formulated as
\begin{subequations}\label{vgr11}
\begin{align}
    \!\!&\!(\hbH_1, \hbH_2) && =  \argmin_{\bbH_1\in \! \ccalH_1,\bbH_2\in \! \ccalH_2} \ \| \bbX \!-\! \bbH_1\bbX \!-\! \bbH_2\bbY \!-\! \bbV\|_F^2 \label{vgr11_a_obj} \nonumber \\
    \!\!&\! && + \alpha\|\bbH_1\|_1 + \beta \|\bbH_2\|_1  \\
    \!\!&\! \mathrm{\;\;s. \;t. } &&  \bbH_2[k,(i,j)] \leq \theta \mathds{1}(\bbH_1[k,i]\bbH_1[k,j] \bbH_1[i,j]); \label{vgr11_b_const}
\end{align}
\end{subequations}
where the second and third terms in \eqref{vgr11_a_obj} (with $\alpha>0$ and $\beta>0$ being hyperparameters) regulate the desired sparsity level in $\bbH_1$ and $\bbH_2$, respectively. By employing $\ccalH_1 = \{ \bbH_1 \geq \bb0, \bbB_1 \circ \bbH_1 = \bb0\}$ and $\ccalH_2 = \{ \bbH_2 \geq \bb0, \bbB_2 \circ \bbH_2 = \bb0\}$ we adopt the structural requirements outlined at the end of Sec. \ref{S:netw_reconstruction} which impose that both $\bbH_1$ and $\bbH_2$ must possess positive weights and no self-loops. The indicator function from the constraint is defined as $\mathds{1}(z)=1$ if $z \neq 0$ and $\mathds{1}(z)=0$ if $z = 0$.
By setting $z = \bbH_1[k,i]\bbH_1[k,j] \bbH_1[i,j]$, we have that if nodes $k$, $i$ and $j$ form a triangle ($z \neq 0$) then the nonlinear relation captured by the Volterra kernel $\bbH_2$ could exist ($\bbH_2[k,(i,j)] \leq \theta \mathds{1}(z)$). If needed, the parameter $\theta$ can be selected to limit the maximum value attributed to each node-pair interaction. Note that having one entry of $\bbH_2$ different from zero implies that the associated triangle is filled.
\vspace{-0.3cm}
\begin{remark}
    Consider that when a triangle exists and is filled (say triangle $(k,i,j)$), the formulation in \eqref{vgr11} does not impose that the values of all node-pair interactions between the nodes present in a filled triangle are the same. If this is required, it can be accomplished by adding the set of constraints $\bbH_2[k,(i,j)]=\bbH_2[k,(j,i)]=\bbH_2[i,(k,j)]=\bbH_2[i,(j,k)]=\bbH_2[j,(k,i)]=\bbH_2[j,(i,k)]$ for all $(k,i,j)$.
\end{remark}
\vspace{-0.3cm}


Problem \eqref{vgr11} is non-convex due to the constraint \eqref{vgr11_b_const}. To deal with it, we apply a group sparsity term that groups all the entries in $\bbH_1$ and $\bbH_2$ that participate in  \eqref{vgr11_b_const}. To that end, we build the $N\times (N+N^2)$ matrix $[\bbH_1, \bbH_2]$ and, since each constraint in \eqref{vgr11_b_const} involves 3 nodes, we construct the $N\times (N+N^2)$ binary mask matrix  $\bbQ^{(i,j,k)}$. This matrix identifies the entries of $[\bbH_1,\bbH_2]$ associated with i) edges between the three nodes $\! \bbQ^{(i,j,k)}[i,\!j] \!\!=\!\! 1$,$\bbQ^{(i,j,k)}[i,\!k]\!\!=\!\! 1$,$\bbQ^{(i,j,k)}[j,\!k]\!\!=\!\!1$, and ii) node-pair interactions between the three nodes $\bbQ^{(i,j,k)}[i,Nj+k] = 1$, $\bbQ^{(i,j,k)}[j,Ni+k] = 1$, $\bbQ^{(i,j,k)}[k,Ni+j] = 1$, with all other entries being zero. 
Leveraging $\bbQ^{(i,j,k)}$, we propose the following convex formulation 

\vspace{-0.4cm}
\begin{alignat}{2}\label{vgr112}
    \!\!&\!(\hbH_1 , \hbH_2)  = &&\!\!\! \argmin_{\bbH_1 \in \ccalH_1, \bbH_2 \in \ccalH_2}  \| \bbX - \bbH_1\bbX - \bbH_2\bbY - \bbV\|_F^2 + \alpha\|\bbH_1\|_1  \nonumber \\
    \!\!&\! &&  + \beta\|\bbH_2\|_1 +\gamma \sum_{i,j,k=1}^{N}\|\bbQ^{(i,j,k)}\circ [\bbH_1, \bbH_2]\|_F, 
\end{alignat}
with $\gamma>0$. In \eqref{vgr112}, we kept the least square term and the sparsity-promoting terms in \eqref{vgr11_a_obj} but replaced \eqref{vgr11_b_const} with a group sparsity term. 
The group-sparsity regularizer links the penalty of activating an entry of $\bbH_1$ with that of activating an entry of $\bbH_2$ provided that those entries can participate in a potential triangle.  When all the interactions (entries) inside the group sparsity norm hold non-zero values, it signifies the presence of a filled triangle.

Although convexity guarantees that problem \eqref{vgr112} can be solved in polynomial time, the required computational complexity is not negligible, especially for medium/large size graphs. The number of terms in the group sparsity constraint scales as $O(N^2)$, and the number of variables and constraints scales as $O(N^3)$.  While this multiplicative growth in the number of variables to optimize is somehow unavoidable when dealing with the design of nontrivial schemes to estimate \emph{high-order interactions}, it emphasizes the importance of developing tailored optimization algorithms that, by exploiting the structure in \eqref{vgr112}, lead to a reduced complexity. We leave this task as future work.

\begin{figure*}
	\centering
	
	\begin{minipage}[c]{.33\textwidth} 
		\includegraphics[width=\textwidth]{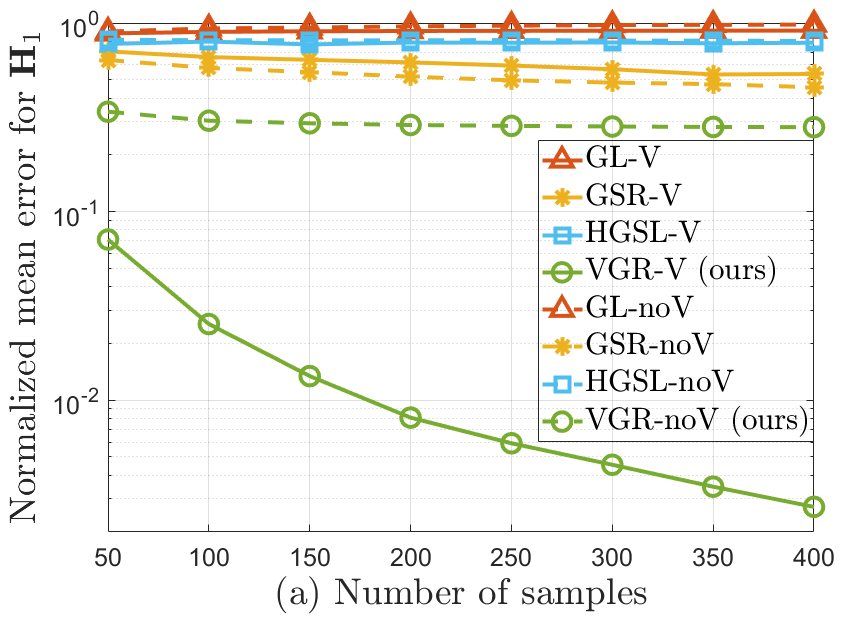}
	\end{minipage}%
	\begin{minipage}[c]{.33\textwidth} 
		\includegraphics[width=\textwidth]{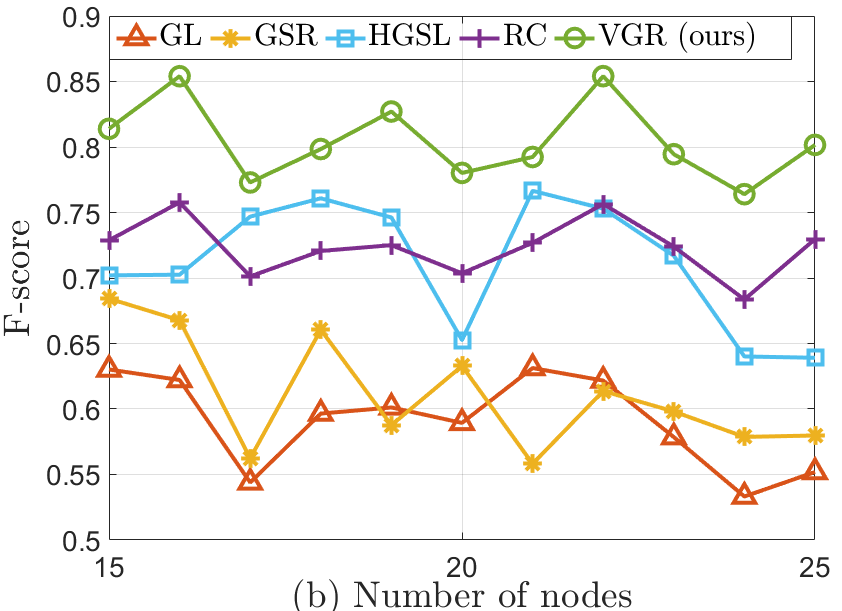}
	\end{minipage}%
	\begin{minipage}{.33\textwidth} 
		\includegraphics[width=\textwidth]{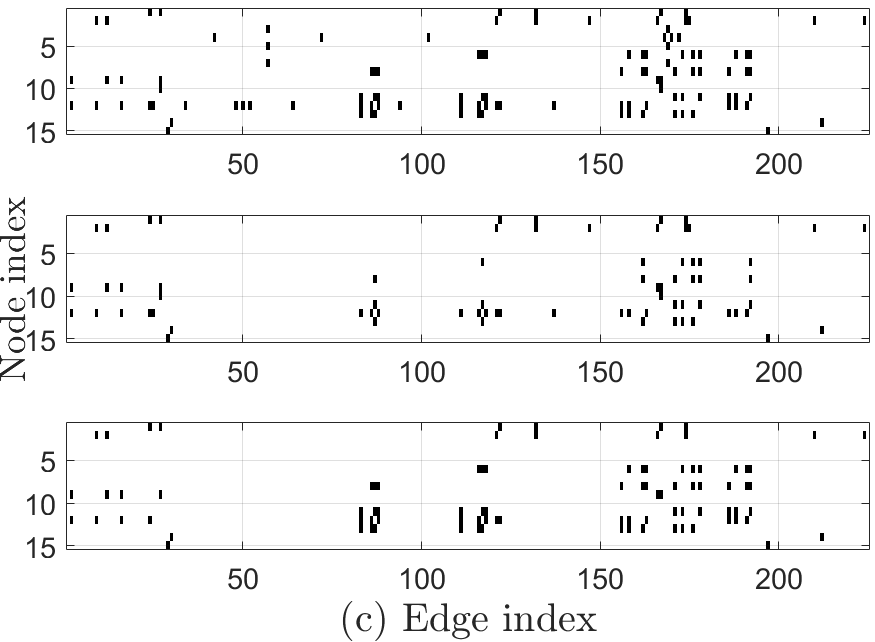}
	\end{minipage}
        \vspace{-0.2cm}
	\caption{(a) Evaluating the estimation performance of different algorithms in terms of normalized squared Frobenius norm [$err(\bbH_1)$, cf. \eqref{nfronorm}] averaged over 50 graph realizations as the number of samples $R$ increases. (b) Evaluating the estimation performance of different algorithms in terms of F-score averaged over 20 graph realizations. (c) Visual representation of the ground-truth support of $\bbH_2$ (upper) and the estimation obtained by RC (middle) and VGR (bottom) for a real-data graph realization with $N=15$ nodes.}
	
	\label{F:exp_12}
\end{figure*}

\section{Numerical experiments}\label{S:numerical_experiments}
\vspace{-0.3cm}

We conduct numerical experiments on both synthetic and real data and compare the following methods.

\begin{itemize}[leftmargin=*]
    \item  GL: Graphical Lasso \cite{GLasso2008}, which learns the edges of a graph by estimating a sparse precision matrix from Gaussian graph signals.
    \vspace{-0.45cm}
    \item GSR: Approach in \cite{segarra2017network}, which estimates the graph topology by assuming the signals are graph stationary in the sought graph.
    \vspace{-0.1cm}
    \item HGSL: Approach in \cite{dualsmooth}, which estimates the graph by assuming smoothness on node (0-simplex) and edge (1-simplex) signals.
    \vspace{-0.1cm}
    \item RC: Rips complex  \cite{zomorodian2010fast}, which estimates SCs (edges and filled triangles) from the correlation of the data.
    \vspace{-0.1cm}
    \item MTV-SC: Approach in \cite{sc_barbarosa}, which estimates SCs from edge signals assuming the topology of the underlying graph is given. 
    \vspace{-0.1cm}
    \item VGR: Our approach in \eqref{vgr112} for estimating the graph and SC (edges and filled triangles) from data using a Volterra signal model.
\end{itemize}

The exact implementation details of the previous schemes and ensuing setups can be found in the online code repository \footnote{https://github.com/andreibuciulea/Graph-SCs-topoID}.

\vspace{0.05cm}
\noindent \textbf{Number of samples.}
For this experiment, we assess the performance of VGR by using the following synthetic data generation setup. 
We generate $M$ graph signals following the autoregressive Volterra model in \eqref{E:x1model2}. The entries of $\bbH_1$ are set as the adjacency matrix of an Erd\H{o}s Rényi graph with $N=20$ nodes and edge probability $p=0.15$ \cite{bollobas1998random}. The entries of $\bbH_2$ are set so that all triangles are filled, which is a favorable setup for the RC algorithm.
Fig. \ref{F:exp_12}.a shows the estimation error for $\bbH_1$ (y-axis) averaged over 50 graph realizations while increasing the number of observed signals $R$ (x-axis). The metric used to compute the estimation error is the normalized squared Frobenius norm: 
\begin{alignat}{2}\label{nfronorm}
    {\rm err}(\bbH_1) = {\| \bbH_1^*-\hbH_1 \|_F^2}/{\|\bbH_1^*\|_F^2}
\end{alignat}
where $\bbH_1^*$ and $\hbH_1$ stand for the ground truth and estimated $\bbH_1$ respectively. Solid lines in Fig. \ref{F:exp_12}.a consider $\bbV$ in $\eqref{E:x1model2}$ known, whereas dashed lines consider $\bbV$ unknown and set to zero.

Starting first with the solid lines, we observe that: i) our algorithm yields the best performance and ii) as $R$ increases, ${\rm err}(\bbH_1)$ decreases. The only exception to ii) is GL, probably because its modeling assumptions are too simple for the signal structure postulated in $\eqref{E:x1model2}$.
When the exogenous variable is unknown (dashed lines), the performance of all algorithms deteriorates. Our approach outperforms the alternatives, due to the consideration of a more comprehensive model that accounts for signal nonlinearities and higher-order interactions. Results for RC were not shown because, in all tested cases, the normalized error was slightly above 1.0. 

\begin{table}[]
\caption{Normalized error when estimating 2-simplices ${\rm err}(\bbH_2)$ for different algorithms and number of samples $R$.}
\centering
\vspace{-0.2cm}
\begin{tabular}{lllllll}
\hline \hline
Alg. $\backslash$ $R$       & \;\; 50 & \;\; 100 & \;\;  200 & \;\;  300 & \;\;  400 & \;\;  500    \\ \hline \hline
\textbf{MTV-SC}      & 1.505 & \;\; 1.496 & \;\; 1.497 & \;\; 1.493 & \;\; 1.494  & \;\; 1.490 \\ 
\textbf{RC}          & 0.790 & \;\; 0.767 & \;\; 0.761 & \;\; 0.753 & \;\; 0.748  & \;\; 0.751\\ 
\textbf{VGR}         & 0.559 & \;\; 0.428 & \;\; 0.294 & \;\; 0.214 & \;\; 0.165  & \;\; 0.133\\ \hline
\end{tabular}
\vspace{-0.1cm}
\label{T:1}
\end{table}

We now move our analysis to assess the performance when estimating $\bbH_2$. 
Since GL and GSR do not consider high-order interactions explicitly, a direct comparison with these methods is infeasible. Thus, we compare the proposed algorithm with MTV-SC and RC.
The results in Table~\ref{T:1} reveal ${\rm err}(\bbH_2)$ is larger than ${\rm err}(\bbH_1)$ as illustrated in Fig. \ref{F:exp_12}.a. This behaviour aligns with expectations, since the task of estimating higher-order interactions is inherently more difficult than estimating links between nodes. This difficulty is due to the large number of potential interactions to estimate compared to the number of available signals.
Nevertheless, our approach not only achieves lower errors than considered alternatives but also exhibits a faster error reduction as the number of samples $R$ increases. We attribute this enhanced $\bbH_2$ estimate to the ability of our approach to simultaneously compute both SCs and the underlying edges in the graph instead of the two-step estimation process implemented by RC and MTV-SC.



\vspace{1mm}
\noindent \textbf{Co-authorship datasets.}
We now evaluate the performance of VGR using a real dataset, following the setup in \cite{dualsmooth}. 
The dataset comprises papers from the ACM conference, featuring 17,431 authors, 122,499 papers, and 1,903 keywords.
The nodes (0-simplices) are a subset of the authors. To establish the ground truth $\bbH_1$, we examined author-paper relationships and considered a link between two authors if they collaborated on a paper.
For ground truth $\bbH_2,$ we considered a filled triangle whenever three authors collaborated on a paper. 
To generate the input signals, we set $R=1,903$ (the total number of different keywords). As a result, the value of each input signal (columns of $\bbX$) is related to the frequency at which a particular author uses a particular keyword across papers.
We constructed 20 different graphs from the dataset, varying the set of authors (with cardinalities between $15$ and $25$), keeping the number of signals as $R=1,903$.  
These generated signals were employed to estimate the graph topology using the different algorithms.
Fig. \ref{F:exp_12}.b displays the average results across 20 graph realizations, represented in terms of F-score (y-axis), with the number of nodes (authors)  ranging from $15$ to $25$ (x-axis).
The results indicate that approaches that do not consider higher-order interactions, such as GL and GSR, yield poorer graph estimations compared to other alternatives. VGR consistently outperforms all other schemes, achieving the highest averaged F-score across all considered graph sizes.

\begin{table}[]
\caption{F-score and ${\rm err}(\bbH_2)$ when estimating 2-simplices from real-data for different algorithms and number of nodes $N$.}
\centering
\vspace{-0.2cm}
\begin{tabular}{l|lll|lll}  
\hline \hline
\multicolumn{4}{c|}{F-score} & \multicolumn{3}{c}{Error} \\
\hline
Alg. $\backslash$ $N$ & \;\; 15 & \;\; 20 & \;\; 25 & \;\; 15 & \;\; 20 & \;\; 25    \\ \hline \hline 
\textbf{MTV-SC}       & \;\; 0.093 & \;\; 0.058 & \;\; 0.056 & \;\; 7.418 & \;\; 7.536 & \;\; 7.530 \\
\textbf{RC}           & \;\; 0.667 & \;\; 0.650 & \;\; 0.585 & \;\; 1.350 & \;\; 2.101 & \;\; 2.837\\ 
\textbf{VGR}          & \;\; 0.718 & \;\; 0.676 & \;\; 0.625 & \;\; 0.548 & \;\; 0.558 & \;\; 0.649\\ \hline
\end{tabular}
\vspace{-0.1cm}
\label{T:2}
\end{table}

Regarding the estimation of the higher-order connectivities shown in Fig. \ref{F:exp_12}.c, RC achieves an F-score of $0.77$, whereas VGR achieves an F-score of $0.88$ (recall that these are the only algorithms that explicitly account for 2-simplices/triplets). Fig.\ref{F:exp_12}.c. further shows the support of $\bbH_2$ for both the RC and VGR  alongside the ground truth. The recovered $\bbH_2$ exhibits a similar structure compared to the ground truth. This implies that both RC and our algorithm effectively estimate the presence of filled triangles, representing interactions between three or more authors collaborating on the same paper. The proposed method also provides better estimations of the associated weights. In the specific realization shown in Fig. \ref{F:exp_12}.c, our method achieves an ${\rm err}(\bbH_2)$ of 0.054, whereas RC yields an error of 1.334. 
Additional results showing the average F-score and ${\rm err}(\bbH_2)$ results for VGR and RC at different numbers of nodes are provided in Table \ref{T:2}. These results reinforce the conclusions drawn from Fig. \ref{F:exp_12}.c. Lastly, we incorporated the results of SC estimation obtained by the MVT-SC approach. For the SC estimation, we used $\bbY$ as edge signals and the graph estimated by the approach presented in \cite{chepuri2017learning}. From the results shown in Table \ref{T:2}, we can conclude that relying solely on the node signals results in an inadequate estimation of SCs for MVT-SC, requiring having access to the actual edge signals.


\vspace{-0.2cm}
\vspace{-0.2cm}

\section{Conclusions}\label{S:conclusions}
\vspace{-0.2cm}
This paper proposed a method to jointly estimate the graph topology (pairwise interactions) and higher-order dependencies (triples) from nodal data by assuming the latter follows a second-order autoregressive graph Volterra model. We incorporated simplicial complex constraints in the said model to estimate sparse-filled triangles as a proxy for triplet interactions. To assess the estimation performance of the proposed algorithm, we conducted experiments on both synthetic and real datasets, revealing consistently superior results compared to those achieved by alternative methods.



\vfill\pagebreak
\bibliographystyle{IEEE}
\bibliography{citations}

\end{document}